\documentclass[prl,preprintnumbers,superscriptaddress,amsmath,amssymb,nofootinbib,twocolumn,floatfix,longbibliography]{revtex4-1}

\usepackage{setspace}
\usepackage{graphicx}
\usepackage{amsmath}  
\usepackage{ulem}

\usepackage{hyperref}
\usepackage{booktabs}    
\usepackage{siunitx}     
\usepackage{nameref}
\usepackage{multirow}
\usepackage{xcolor}

\newcommand{\sectionname}[1]{ \noindent{\bfseries #1}.---}

\newcommand{\be}{\begin{equation}}
\newcommand{\ee}{\end{equation}}

\newcommand{\dd}{\mathrm{d}}

\usepackage{soul}

\begin{document}

\preprint{KCL-PH-TH-2026-17}
 
\title{Detector orientation matters:
Consequences for gravitational-wave background detectability}

\author{Nelson Christensen}
 \email{nelson.christensen@oca.eu}
 \affiliation{
 Université Côte d’Azur, Observatoire de la Côte d’Azur, CNRS, Laboratoire Artemis, 06300 Nice, France}

\author{Joseph D.Romano}
\email{joseph.romano@utrgv.edu}
\affiliation{Department of Physics and Astronomy, University of Texas Rio Grande Valley, Brownsville TX 78520, USA}

\author{Mairi Sakellariadou}
\email{mairi.sakellariadou@kcl.ac.uk}
\affiliation{Theoretical Particle Physics and Cosmology Group,  Physics Department, \\ King's College London, University of London, Strand, London WC2R 2LS, United Kingdom}
\affiliation{
 Université Côte d’Azur, Observatoire de la Côte d’Azur, CNRS, Laboratoire Artemis, 06300 Nice, France}

\author{Jishnu Suresh}
 \email{jishnu.suresh@oca.eu}
 \affiliation{
 Université Côte d’Azur, Observatoire de la Côte d’Azur, CNRS, Laboratoire Artemis, 06300 Nice, France}

\date{\today}

\begin{abstract}
Cross-correlation searches for gravitational-wave backgrounds depend on the geometrical configuration (physical separation and relative orientation) of the detectors comprising the network.
Applying standard techniques to a few simple examples, we illustrate how the relative orientation of a pair of Earth-based L-shaped laser interferometers can drastically impact the detectability of both isotropic and anisotropic gravitational-wave backgrounds.
\end{abstract}

\maketitle

\sectionname{Introduction}
\label{Sec:Intro}
The detection of the gravitational-wave background (GWB) formed by the superposition of gravitational waves (GWs) from astrophysical and cosmological sources will provide essential information for the astrophysics of compact objects~\cite{LIGOScientific:2025bgj}, as well as for high-energy physics and early-universe cosmological models at energy scales otherwise unattainable~\cite{Christensen:2018iqi,LIGOScientific:2025kry}. Given the importance of detecting a GWB, we illustrate how the relative orientation of two arbitrary terrestrial L-shaped interferometers may affect the detectability of a GWB for both isotropic and anisotropic distributions of GW power on the sky. 

The techniques we use are standard in the field~\cite{10.1093/mnras/227.4.933,PhysRevD.46.5250,PhysRevD.48.2389,PhysRevD.55.448,Romano:2016dpx}. We gather these results together in this short presentation, in light of decisions that are being made for the location and orientation of the next generation of ground-based detectors. We follow a general approach in our study, in order for our conclusions to be applicable to any foreseen future interferometric detector network.
\medskip

\sectionname{Assumptions and geometrical configuration}
\label{s:assumptions}
For concreteness, we will consider two L-shaped laser interferometers, indexed by 1 and 2, each having $L\equiv 10~{\rm km}$-long arms, and located at positions $\vec x_1$ and $\vec x_2$ on the surface of the Earth.
We will assume that the two interferometers are separated by a distance $d\equiv 1000~{\rm km}$ as measured along the arc of the great circle (a geodesic) connecting the vertices of the L's.%
\footnote{The values of $L$ and $d$ that we chose are purely representative (i.e, order-of-magnitude) values.
We can make them larger or smaller by a factor of a few without significantly changing the final results.}
The relative orientation of the two detectors is defined as the difference between the angles that their respective $\hat u$ arms make with respect to the great circle connecting the two detectors.
The $\hat v$ arms of each of the two interferometers make an angle of $90^\circ$ with respect to their $\hat u$ arms, given our assumption of L-shaped interferometers.

\begin{figure}
\centering
\includegraphics[width=0.45\textwidth]{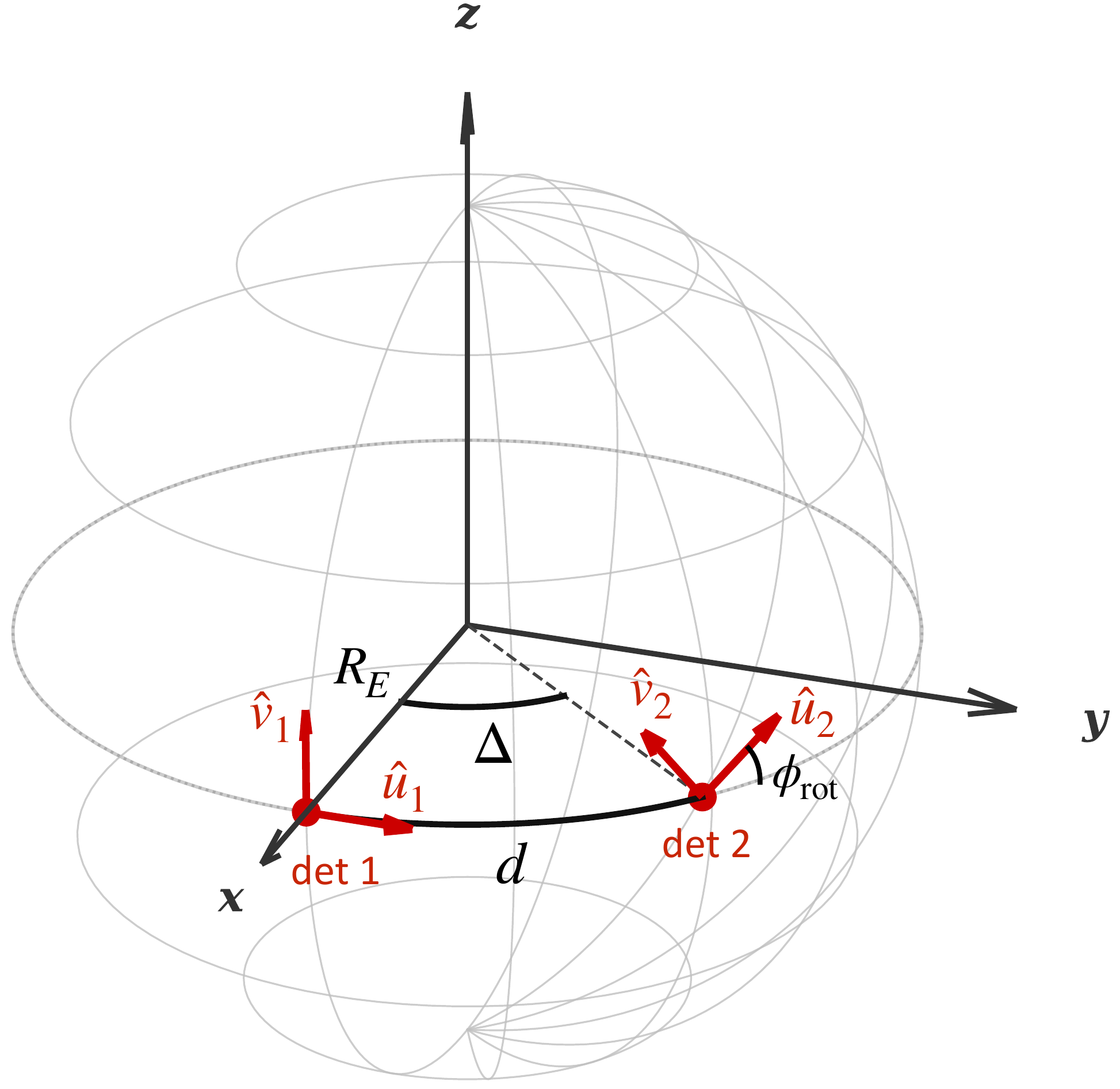}
\caption{Geometrical configuration of the two L-shaped interferometers.}
\label{f:geometry}
\end{figure}

Fig.~\ref{f:geometry} shows in detail the geometrical configuration of the interferometers that we use for later calculations.
For simplicity, we will assume that the Earth is a perfect sphere with radius $R_E = 6400~{\rm km}$, and we choose a rectangular coordinate system with the $z$-direction perpendicular to the plane defined by the great circle passing through the vertices of the two interferometers.
We orient the $x$ and $y$ axis so that the $x$-axis passes through the vertex of detector 1.
The position vectors of the vertices of the two interferometers are then
\be
\vec x_1 = R_E\,\hat x\,,
\quad
\vec x_2 = R_E( \cos\Delta\, \hat x + \sin\Delta\,\hat y)\,,
\label{e:x1x2}
\ee
where $\Delta \equiv d/R_E$. 
For simplicity, we take the $\hat u$ arm of detector 1 to point along the great circle in the direction of detector 2 (with the $\hat v$ arm perpendicular to $\hat u$), so that 
\be
\hat u_1=\hat y\,,\qquad
\hat v_1=\hat z\,.
\label{e:u1v1}
\ee
The corresponding unit vectors for the arms of detector 2 are defined by first parallel-propagating $\hat u_1$ and $\hat v_1$ to $\vec x_2$ along the great circle, and then rotating these vectors counter-clockwise (about $\vec x_2$) by $\phi_{\rm rot}$.
This leads to 
\be
\begin{aligned}
&\hat u_2=
+\cos\phi_{\rm rot}\,(-\sin\Delta\, \hat x +\cos\Delta\,\hat y) + \sin\phi_{\rm rot}\, \hat z\,,
\\
&\hat v_2=
-\sin\phi_{\rm rot}\,(-\sin\Delta\, \hat x +\cos\Delta\,\hat y) + \cos\phi_{\rm rot}\, \hat z\,.
\label{e:u2v2}
\end{aligned}
\ee
Note that $\phi_{\rm rot}=0$ ($\phi_{\rm rot}=45^\circ$) corresponds to the maximally-aligned (maximally-misaligned) configuration, taking into account the curvature of the Earth.
The two LIGO interferometers have $\phi_{\rm rot}=90^\circ$, which is equivalent to $\phi_{\rm rot}=0$ for the GWB search.
Equations~\eqref{e:x1x2}, \eqref{e:u1v1}, \eqref{e:u2v2} completely define the geometrical configuration of our two interferometers.

\begin{figure}[h]
\centering
\includegraphics[width=0.45\textwidth]{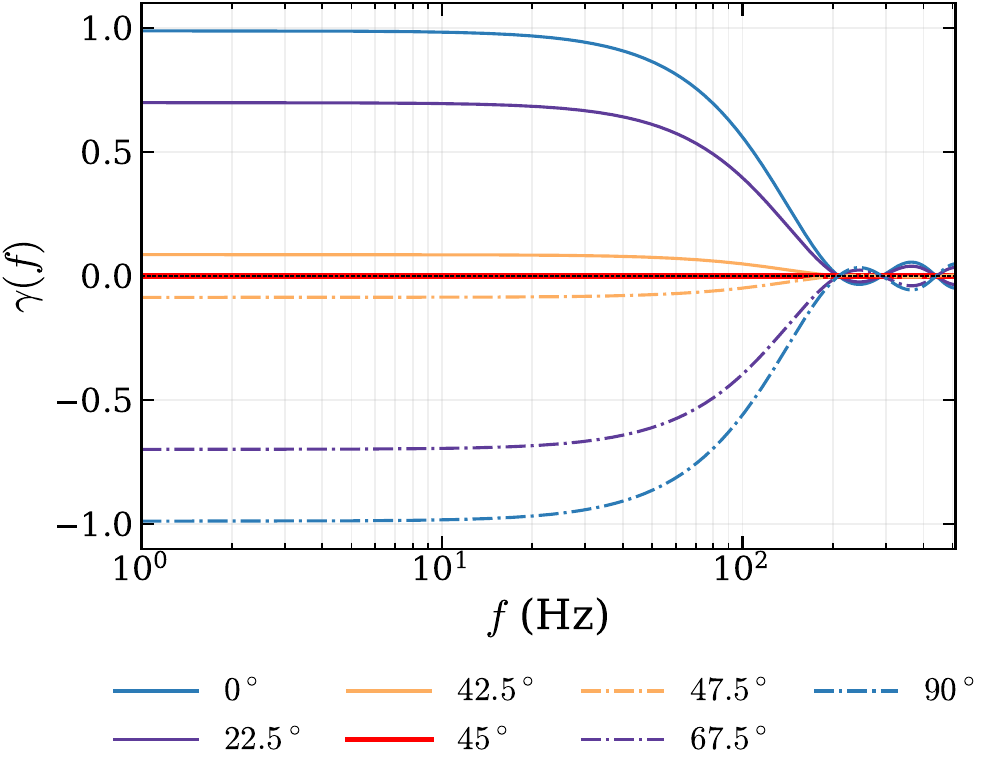}
\caption{Overlap reduction functions plotted as a function of frequency for several orientation angles $\phi_{\rm {rot}}$ between the two interferometers.}
\label{f:orf_vs_angle}
\end{figure}

We will also assume that we are in the {\it short-antenna limit}, where the wavelengths $\lambda\equiv c/f$ of the GWs are assumed to be much larger than the arm lengths $L$ of the interferometers. 
This is a valid approximation for 10-km long arms and GW frequencies of order $f\lesssim 1000~{\rm Hz}$  (so wavelengths $\lambda\gtrsim 300~{\rm km}$).  
But note that the physical separation $s\equiv |\vec x_1-\vec x_2|$ of the two detectors need not be small compared to the GW wavelengths, so factors of $s/\lambda$ are not necessarily $\ll 1$. 
(Hence our choice of {\it short-antenna limit} is more informative than the often-used {\it long-wavelength approximation} terminology.)
\footnote{Using the exact detector response functions would change e.g., the curves in Fig.~\ref{f:orf_vs_angle}
by introducing ``wiggles" starting at about 100 Hz and having absolute values less than a few times $10^{-3}$.}

Finally, for the signal-to-noise ratio (SNR) calculations that we perform below, we will assume that the two interferometers have the same noise power spectral densities (units of ${\rm strain}^2/{\rm Hz}$), which we take to be constant between $f_{\rm low}=10~{\rm Hz}$ and $f_{\rm high}=100~{\rm Hz}$.
In addition, we will assume that the amplitude of the detector noise is much greater than that of the GWB for all frequencies [i.e., $P_{n_1}(f)\gg P_{\rm gw}(f)$], allowing us to work with expressions that are valid in the {\it weak-signal limit}.
\medskip

\sectionname{Isotropic GWBs}
\label{Sec:Iso}
The key quantity that affects searches for isotropic GWBs is the {\it overlap reduction function} (ORF) $\gamma(f)$ between the two detectors.%
\footnote{Since we are not considering multiple detector pairs in this note, we do not need to put $12$ indices on $\gamma(f)$ to distinguish it from the ORFs for other detector pairs.
This allows us to simplify some of the expressions involving the single pair of detectors.}
For the two interferometers described above, the ORF is defined by~\cite{10.1093/mnras/227.4.933,PhysRevD.46.5250,PhysRevD.48.2389,PhysRevD.55.448,Romano:2016dpx}
\begin{equation}
\gamma(f) \equiv \frac{5}{8\pi}
\int {\rm d}^2\Omega_{\hat k}\sum_{A=+,\times} F_1^A(\hat k)F_2^A(\hat k) e^{i2\pi f\hat k\cdot(\vec x_1-\vec x_2)/c}\,,
\label{e:gammaIJ}
\end{equation}
where the normalization has been chosen so that $\gamma(f)=1$ for two colocated and coaligned L-shaped interferometers in the short-antenna limit.
In addition,
\be
F_I^A(\hat k) \equiv e^A_{ij}(\hat k) D_I^{ij}\,,\quad
D_I^{ij}\equiv \frac{1}{2}\left(\hat u_I^i\hat u_I^j-  \hat v_I^i\hat v_I^j\right)
\label{e:FandD}
\ee
are the individual antenna response functions and detector tensors for detector $I=1,2$ (with repeated spatial indices $i$, $j$ summed over).
In the above expressions, $\hat k$ is the propagation direction of a plane-polarized GW having frequency $f$ and corresponding polarization tensor $e_{ij}^A(\hat k)$.
The antenna response functions are calculated in the short-antenna limit discussed earlier, and hence are independent of frequency.

As shown in \cite{Allen:1997ad}, it is also possible to write down an analytic expression for the ORF for our two interferometers in the short-antenna limit as a sum of three terms involving contractions of the detector tensors $D_1^{ij}$, $D_2^{ij}$, and the unit separation vector $\hat s \equiv (\vec x_1-\vec x_2)/s$, multiplied by functions of $\alpha \equiv 2\pi fs/c$.
Explicitly,
\be
\begin{aligned}
\gamma(f) &= A(\alpha)D_1^{ij}D_2^{ij}+ B(\alpha)D_1^{ij} D_2^{ik}\hat s_j\hat s_k
\\
&\hspace{0.2in}
+C(\alpha)D_1^{ij}D_2^{kl}\hat s_i \hat s_j \hat s_k \hat s_l\,,
\label{e:gamma_analytic}
\end{aligned}
\ee
where
\be
\begin{bmatrix}
A(\alpha) \\
B(\alpha) \\
C(\alpha)
\end{bmatrix}
=
\frac{1}{2\alpha^2}
\begin{bmatrix}
10\alpha^2 & -20\alpha & 10 \\
-20\alpha^2 & 80\alpha & -100 \\
5\alpha^2 & -50\alpha & 175
\end{bmatrix}
\begin{bmatrix}
j_0(\alpha) \\
j_1(\alpha) \\
j_2(\alpha)
\end{bmatrix}\,,
\ee
with $j_0(\alpha)$, $j_1(\alpha)$, $j_2(\alpha)$ being the first three spherical Bessel functions.
This analytic expression for the overlap reduction function is particularly convenient for showing that for the maximally-misaligned case (i.e., for $\phi_{\rm rot}=45^\circ)$, $\gamma(f)=0$ for all frequencies regardless of the non-zero physical separation of the detectors.
This result is a consequence of the short-antenna limit and having the $\hat u$-arm of one detector point along the geodesic connecting the vertices of the two interferometers \cite{SuppMaterial}.

Figure~\ref{f:orf_vs_angle} shows the overlap reduction functions  plotted as a function of frequency for several orientation angles $\phi_{\rm rot}$ between $0^\circ$ and $180^\circ$. The first zero in the overlap reduction occurs at $f \approx c/2s$. Increasing the distance between the detectors has the effect of reducing the sensitivity to a GWB search. With a global network of detectors it would be important to align ($\phi_{\rm rot} \approx 0$) the two detectors with the smallest distance separation.

For isotropic GWBs, we can also calculate the expected SNR for the standard cross-correlation search.
Recall that the energy density spectrum for GWs is defined by
\be
\Omega_{\rm gw}(f) \equiv\frac{1}{\rho_{\rm crit}}\frac{\dd \rho_{\rm gw}}{\dd \ln f} 
\ee
where $\rho_{\rm crit} \equiv 3 H_o^2 c^2/8\pi G$ is the critical energy density needed today to close the universe.
It is related to the strain spectral density of the GWB via
\be
S_h(f) \equiv\frac{3 H_0^2}{2 \pi^2} \frac{\Omega_{\rm gw}(f)}{f^3}\,.
\ee
In terms of these quantities, one can show~\cite{Thrane:2013oya} that the square of the expected cross-correlation SNR $\rho$ is given by
\be
\rho^2 = 2T
\int_{f_{\rm low}}^{f_{\rm high}}
\dd f\>\frac{S_h^2(f)(\gamma(f)/5)^2}
{P_{n_1}(f) P_{n_2}(f)}\,,
\label{e:rho_1so}
\ee
where $T$ is the observation time (which we will take to be one year).

\begin{figure}[h]
\centering
\includegraphics[width=0.45\textwidth]{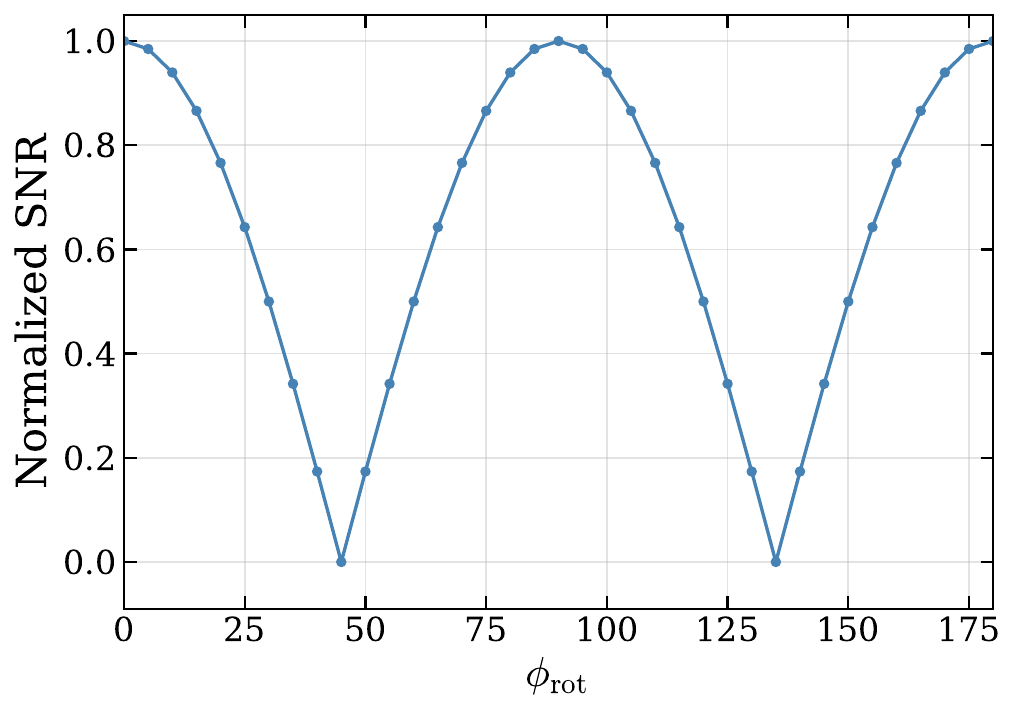}
\caption{Normalized SNR for a cross-correlation search for an isotropic background as a function of the orientation angle $\phi_{\rm rot}$ between the two interferometers, showing peaks at $\phi_{\rm rot} = 0^{\circ}$, $90^{\circ}$, $180^{\circ}$ and exact nulls at $45^{\circ}$ and $135^{\circ}$.}
\label{f:rho_vs_angle}
\end{figure}

Fig.~\ref{f:rho_vs_angle} shows a normalized version of the expected SNR $\rho$ plotted as a function of the orientation angle $\phi_{\rm rot}$ between $0^\circ$ and $180^\circ$. 
The normalized SNR is computed as $\rho_{\rm norm}(\phi_{\rm rot}) = \rho(\phi_{\rm rot})\,/\,\rho_{\rm max}$, with the normalization taken with respect to the optimal orientation $\phi_{\rm rot} = 0^{\circ}$. 
Here we set the noise power spectral densities to constant values and assume the spectral shape of $\Omega_{\rm {gw}}$ to be constant.
One can see from this figure that the SNR peaks at $\phi_{\rm rot} = 0^{\circ}$, $90^{\circ}$, and $180^{\circ}$, where the detectors are aligned, and vanishes exactly at $\phi_{\rm rot} = 45^{\circ}$ and $135^{\circ}$, where the overlap reduction function $\gamma(f) = 0$ for all frequencies.

Consider a detection of $\Omega_{\rm gw} = 10^{-13}$ with ${\rm SNR} = 4$ for the optimal orientation $\phi_{\rm{rot}} = 0^{\circ}$. One can directly deduce the detection level and SNR to any other orientation using the normalized SNR plot. At $\phi_{\rm{rot}}  = 45^{\circ}$, the normalized SNR vanishes ($\rho_{\rm norm} = 0$), meaning that the background becomes completely undetectable regardless of its amplitude. Moving slightly away from this null, at $\phi_{\rm{rot}} = 42.5^{\circ}$ where $\rho_{\rm norm} = 0.0872$, achieving a detection with the same ${\rm SNR} = 4$ would require $\Omega_{\rm gw} \approx 1.15 \times 10^{-12}$, nearly 11 times larger than the optimal case.

\begin{figure*}
\centering
\includegraphics[width=\textwidth]{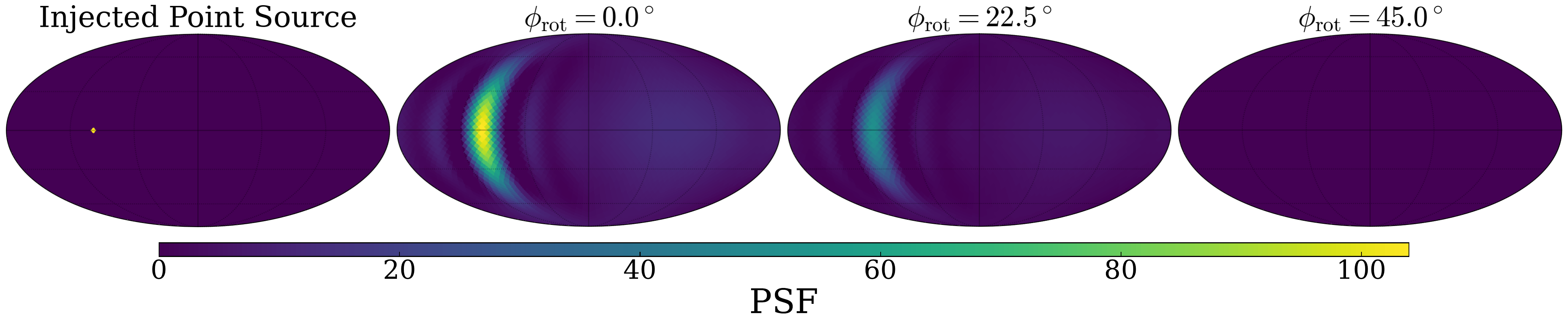}
\caption{Effect of orientation angle on the point spread function for a delta-function point source. The leftmost panel shows the injected source, while the remaining panels display the resulting point spread functions for orientation angles $\phi_{\rm rot} = 0^\circ$, $22.5^\circ$, and $45^\circ$.} 
\label{f:psf_vs_angle_pointsource}
\end{figure*}
\begin{figure*}
\centering
\includegraphics[width=\textwidth]{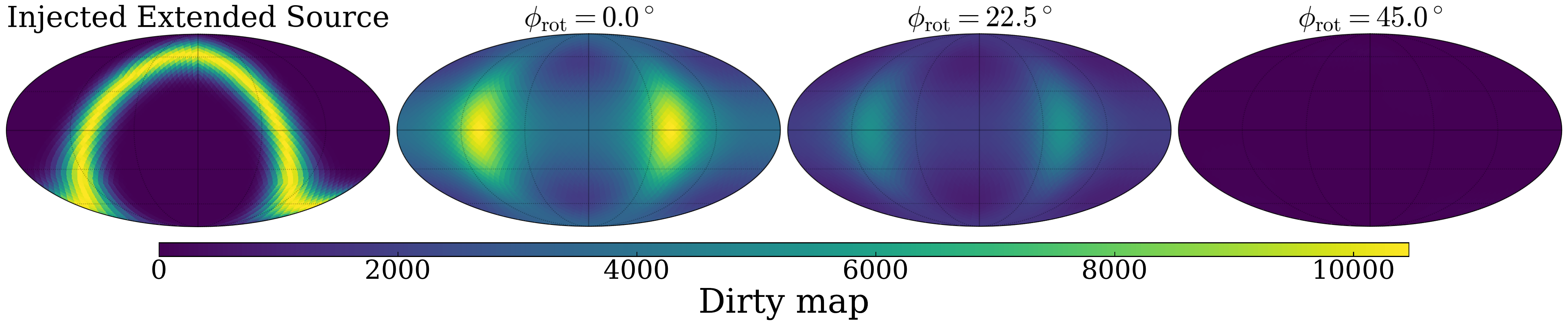}
\caption{Effect of orientation angle on the point spread function for an extended source. The leftmost panel shows the injected source, while the remaining panels display the resulting point spread functions for orientation angles $\phi_{\rm rot} = 0^\circ$, $22.5^\circ$, and $45^\circ$.} 
\label{f:psf_vs_angle_extendedsource}
\end{figure*}

\medskip
\sectionname{Anisotropic GWBs}
\label{Sec:Dir}
The key quantity that affects searches for anisotropic GWBs is (up to an overall normalization factor) the integrand of the ORF:
\be
\gamma(t;f,\hat k)
\equiv 
\frac{1}{2}
\sum_{A=+,\times} F_1^A(\hat k,t)F_2^A(\hat k,t) e^{i2\pi f\hat k\cdot(\vec x_1(t)-\vec x_2(t))/c}\,,
\ee
where we have indicated the time-dependence of the antenna response functions and the position vectors of the two interferometers due to the Earth's rotation about its axis.
This quantity arises when calculating the expected value of the cross-correlation estimator
\be
\hat C(t;f)
\equiv \frac{2}{\tau}
\tilde d_1(t;f)\tilde d_2^*(t;f)\,,
\ee
where $\tilde d_I(t;f)$ is the {\it short-term} Fourier transform of the time-series data from detector $I=1,2$ having duration $\tau$ centered at time $t$.
The duration $\tau$ is taken to be sufficiently short (of order a couple of minutes) in order that the position of the detectors relative to the sky does not change very much due to Earth's rotation.
As shown in \cite{Romano:2016dpx},
\be
\langle \hat C(t;f)\rangle 
= \bar H(f) \int \dd^2\Omega_{\hat k}\> \gamma(t;f,\hat k) {\cal P}(-\hat k)\,,
\label{e:exp_corr}
\ee
where ${\cal P}(-\hat k)$ is the power in the GWB coming from direction $\hat n=-\hat k$ on the sky and $\bar H(f)$ is the spectral shape of the GWB, normalized to unity at some reference frequency $f_{\rm ref}$.  
This means that $\bar H(f)$ is related to $S_h(f)$ via $S_h(f)=S_h(f_{\rm ref})\,\bar H(f)$.

It is then common to search for an anisotropic GWB by trying to estimate ${\cal P}(-\hat k)$ for a finite number of pixels on the sky from the measured cross-correlations $\hat C(t;f)$.
One way to do this, as described in detail in \cite{PhysRevD.80.122002}, is to construct estimators
\be
\hat{\cal P}_{\hat k}
\equiv
\sum_{\hat k'}(F^{-1})_{\hat k \hat k'}
X_{\hat k'}\,,
\label{e:hatP}
\ee
where
\begin{align}
X_{\hat k}
&\equiv
\sum_t\sum_f\gamma^*(t;f,\hat k)\frac{\bar H(f)}{P_{n_1}(t;f) P_{n_2}(t;f)}\hat C(t;f)\,,
\label{e:X}
\\
F_{\hat k\hat k'}
&\equiv
\sum_t\sum_f\gamma^*(t;f,\hat k)\frac{\bar H^2(f)}{P_{n_1}(t;f) P_{n_2}(t;f)}\gamma(t;f,\hat k')
\,.
\label{e:F}
\end{align}
These quantities are the so-called ``dirty map" and Fisher matrix for the anisotropic search.

One measure of the sensitivity of a search for an anisotropic background is the size and shape of the {\it point spread function} induced by the correlated response of a pair of interferometers to an anisotropic distribution of GW power on the sky.
Given the relatively broad angular distribution of the individual antenna response functions, a GW point source on the sky will be ``smeared" by the response.
As evident from the expected value of the inverse of Eq.~\eqref{e:hatP}, i.e.,
\be
\langle X_{\hat k}\rangle = \sum_{\hat k'} F_{\hat k\hat k'}P_{\hat k'}\,,
\ee
the Fisher matrix plays the role of a point spread function with 
\be
{\rm PSF}_{\hat k_0}(\hat k) 
\equiv F_{\hat k \hat k_0}\,.
\ee
Here, one interprets $-\hat k_0$ as the direction on the sky to a true point source and $F_{\hat k \hat k_0}$ as its spreading to other directions $-\hat k$.
In Figs.~\ref{f:psf_vs_angle_pointsource} and ~\ref{f:psf_vs_angle_extendedsource}, we show the effect of different choices of the orientation angle between the two detectors on the point spread function for both a delta-function point source and an extended source on the sky.

Alternatively, if we know in advance the spatial distribution of the GW power on the sky, but not its overall amplitude $A\equiv \Omega_{\rm gw}(f_{\rm ref})$, then we can construct an optimal estimator for $A$ using an approach similar to matched filtering~\cite{PhysRevD.80.122002, Talukder:2010yd,Agarwal:2022lvk} but for a GWB.
An explicit example of this is a targeted search for the kinematic dipole~\cite{Chung:2022xhv}, which arises due to our motion relative to the cosmic rest frame.
\begin{figure}[htbp]
\centering
\includegraphics[width=0.4\textwidth]{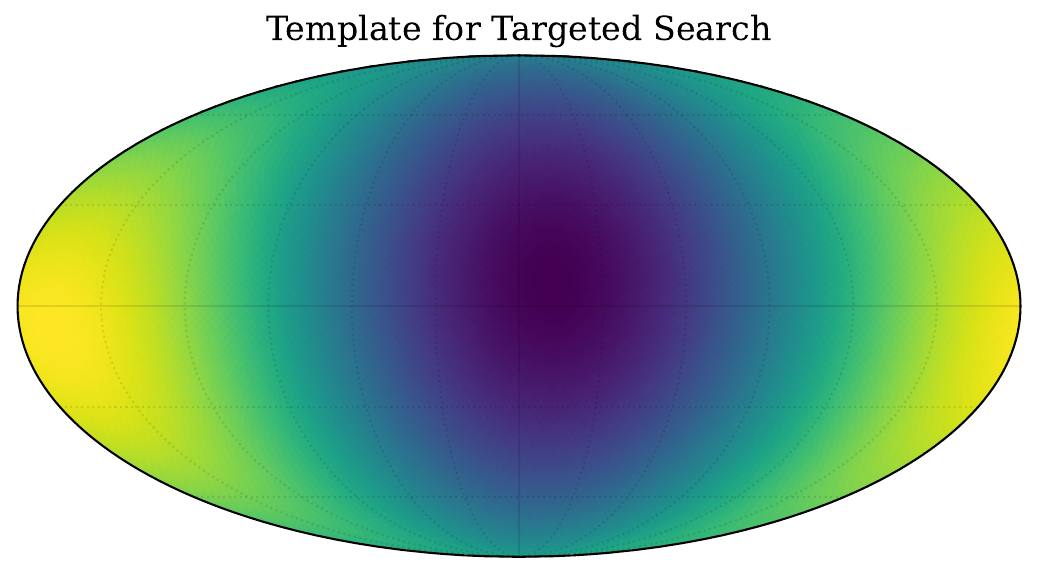}
\caption{Template for the kinematic dipole as determined from CMB measurements. The kinematic dipole arises from the motion of the Solar System with respect to the CMB rest frame at a velocity $v = 369.82$~km/s, corresponding to $\beta \equiv v/c = 1.23 \times 10^{-3}$~\cite{Planck:2018nkj}. This introduces an anisotropic modulation of amplitude $A = \beta\,\Omega_{\rm gw}(f_{\rm ref})$ on top of the isotropic background $\Omega_{\rm gw}(f)$.}
\label{f:dipole_template}
\end{figure}
Using the direction of the dipole as determined from cosmic microwave background (CMB) measurements (as shown in Fig. \ref{f:dipole_template}), we can construct the following estimator of its amplitude:
\be
\hat A = \frac{\sum_{\hat k}\bar{\cal P}_{\hat k}X_{\hat k}}
{\sqrt{\sum_{\hat k, \hat k'}\bar{\cal P}_{\hat k} F_{\hat k\hat k'} \bar{\cal P}_{\hat k'}}}\,,
\ee
where $\bar{\cal P}_{\hat k}$ is the target dipole anisotropy.
The square of the expected SNR of this estimator is then
\be
\rho^2 = A^2\,\sum_{\hat k, \hat k'}\bar{\cal P}_{\hat k} F_{\hat k\hat k'} \bar{\cal P}_{\hat k'}\,.
\label{e:rho_aniso}
\ee
Using Eq.~\eqref{e:F} to substitute for $F_{\hat k\hat k'}$, we can also write
\be
\rho^2 =
\sum_{t,f}\sum_{\hat k, \hat k'}
\frac{A^2 \bar H^2(f)\bar{\cal P}_{\hat k}\,\gamma^*(t;f,\hat k)\gamma(t;f,\hat k')\bar{\cal P}_{\hat k'}}{P_{n_1}(t;f) P_{n_2}(t;f)}\,.
\ee
Note the similarity of this last expression to Eq.~\eqref{e:rho_1so} for the isotropic search.

\begin{figure}[htbp]
\centering
\includegraphics[width=0.45\textwidth]{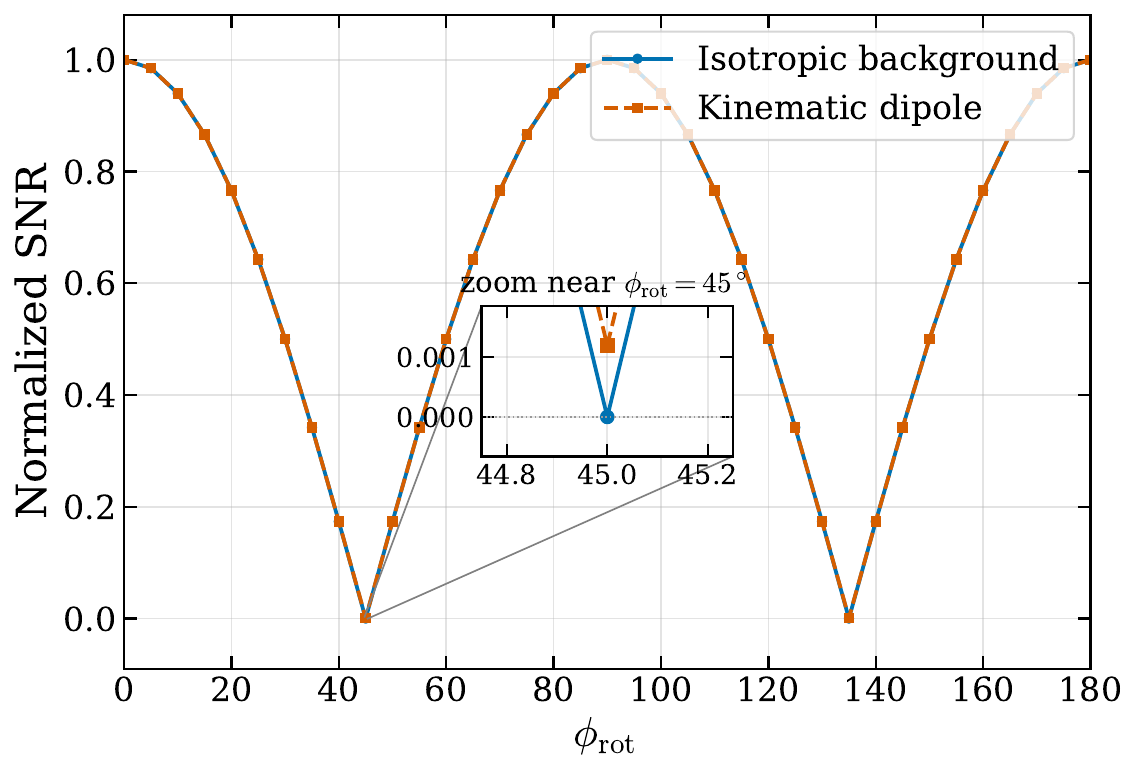}
\caption{Normalized SNR for a targeted kinematic dipole search as a function of the orientation angle $\phi_{\rm rot}$ between the two interferometers, showing peaks at $\phi_{\rm rot} = 0^{\circ}$ and $90^{\circ}$ and a near-null of $\rho_{\rm norm} = 0.0027$ at $\phi_{\rm rot} = 45^{\circ}$. See inset. For comparison we also plot the SNR for the isotropic background.}
\label{f:rho_aniso_vs_angle}
\end{figure}

Fig.~\ref{f:rho_aniso_vs_angle} shows how the expected SNR for the targeted kinematic dipole search varies as a function of the orientation angle between the two interferometers. For this calculation, we set $\bar{H}(f) \propto f^{-3}$ and the noise power spectral densities to constant values, and we normalized the SNR by its value for two equally aligned interferometers---i.e., $\phi_{\rm rot}=0$. Similar to the isotropic case, the SNR peaks at $\phi_{\rm rot} = 0^{\circ}$ and $90^{\circ}$, and only retaining a residual of $\rho_{\rm norm} = 0.0027$ at $\phi_{\rm rot} = 45^{\circ}$.

For an isotropic GWB of $\Omega_{\rm gw} = 10^{-13}$, the kinematic dipole amplitude is $A  
\approx 1.23 \times 10^{-3} \, \Omega_{\rm gw} \approx 1.23 \times 10^{-16}$, since $\rho_{\rm dipole} \propto A \,\rho_{\rm norm}(\phi_{\rm{rot}} )$. At $\phi_{\rm{rot}}  = 45^{\circ}$, where $\rho_{\rm norm} = 0.0027$, the dipole signal is effectively undetectable, while at $\phi_{\rm{rot}}  = 42.5^{\circ}$ where $\rho_{\rm norm} = 0.0872$, the required amplitude to maintain the same SNR scales up by a factor of $\sim 11$ relative to the optimal orientation, reaching $A \approx 1.41 \times 10^{-15}$.
\medskip

\sectionname{Conclusion}
\label{Sec:Conclusions}
As illustrated by the examples presented in this study, detector orientation does matter in searches for a GWB.
We hope these results are considered when decisions regarding the locations and orientations of the next generation of ground-based detectors are  made.

\section*{Acknowledgements} This work was supported by the French government through the France 2030 investment plan managed by the National Research Agency (ANR), as part of the Initiative of Excellence of Université Côte d’Azur under Reference No. ANR-15-IDEX-01.
JDR acknowledges support from (NSF) Grant No. PHY-2207270 and start-up funds at the University of Texas Rio Grande Valley.
The work of MS is partially supported by the Science and Technology Facilities Council (STFC grant ST/X000753/1).

\bibliography{ref}

\clearpage
\onecolumngrid
\section{ SUPPLEMENTAL MATERIAL -- Orientation matters:
Consequences for gravitational-wave background detectability}

\section{Proof that $\gamma(f)=0$ for $\phi_{\rm rot} = 45^\circ$}

To prove that $\gamma(f)$ is identically zero for $\phi_{\rm rot}=45^\circ$, we show that the three geometrical factors that appear in Eq. (6):
\be
D_1^{ij} D_2^{ij}\,,
\quad
D_1^{ij} D_2^{ik}\hat s_j \hat s_k\,,
\qquad
D_1^{ij}D_2^{kl}\hat s_i \hat s_j \hat s_k \hat s_l\,,
\ee
 vanish separately for $\phi_{\rm rot}=45^\circ$.
To see this, we first note that to calculate the detector tensors $D_I^{ij}$ defined by Eq. (5), we need to know $\hat u_1$, $\hat v_1$, $\hat u_2$, $\hat v_2$.
These are given by
\be
\begin{aligned}
&\hat u_1 = \hat y\,,
\\
&\hat v_1 = \hat z\,,
\\
&\hat u_2\propto -\sin\Delta\, \hat x + \cos\Delta\,\hat y + \hat z\,,
\\
&\hat v_2\propto +\sin\Delta\, \hat x - \cos\Delta\,\hat y + \hat z\,,
\label{e:uv_45}
\end{aligned}
\ee
where we used Eqs. (2) and  (3) with $\phi_{\rm rot}=45^\circ$.
Using Eqs.~\eqref{e:uv_45} and (5), it then follows that
\be
\begin{aligned}
D_1^{ij}D_2^{ij} 
&\propto(\hat u_1\cdot\hat u_2)^2 
+(\hat v_1\cdot\hat v_2)^2 
-(\hat u_1\cdot\hat v_2)^2
-(\hat v_1\cdot\hat u_2)^2\\
&=(\cos\Delta)^2 + (1)^2 - (-\cos\Delta)^2 - (1)^2\\
&=0\,.
\end{aligned}
\ee
Next, note that since Eq. (1) implies that $\hat s\equiv (\vec x_1-\vec x_2)/s$ lies in the $xy$ plane, it follows that 
\be
\hat v_1\cdot \hat s=0\,,
\qquad
\hat u_2\cdot \hat s = -\hat v_2\cdot \hat s\,.
\ee
These last two results together with Eq.~\eqref{e:uv_45} imply
\begin{align}
&D_1^{ij} \hat s_j 
\propto \hat u_1(\hat u_1\cdot\hat s) - \hat v_1(\hat v_1\cdot \hat s) \propto \hat u_1=\hat y\,,
\\
&D_2^{ik} \hat s_k
\propto \hat u_2(\hat u_2\cdot\hat s) - \hat v_2(\hat v_2\cdot \hat s) 
\propto \hat u_2+\hat v_2
\propto \hat z\,,
\end{align}
which in turn imply
\begin{align}
&D_1^{ij} D_2^{ik}\hat s_j \hat s_k
\propto \hat y\cdot\hat z = 0\,,
\\
&D_1^{ij} D_2^{kl}\hat s_i\hat s_j \hat s_k\hat s_l
\propto (\hat y\cdot \hat s)(\hat z\cdot\hat s)=0\,.
\end{align}
This completes the proof.

\section{Going beyond the weak-signal approximation}

The following is a back-of-the envelope argument to determine the reduction in SNR if one chooses e.g., $\phi_{\rm rot}=42.5^\circ$ 
(as opposed to $\phi_{\rm rot}=0^\circ$) for the orientation angle between the two L-shaped interferometers.

If we ignore any frequency dependence in the overlap reduction function and the GWB and detector noise power spectral densities, then the expected squared-SNR is proportional to 
\be
\rho^2(\phi_{\rm rot}) = \frac{\gamma^2(\phi_{\rm rot}) \, S^2}{(N+S)^2 + \gamma^2(\phi_{\rm rot}) \, S^2}
= \frac{\gamma^2(\phi_{\rm rot}) \, S^2}{N^2 +2NS + (1 + \gamma^2(\phi_{\rm rot})) \, S^2}\,,
\ee
where we are using $S$ and $N$ to denote the power spectral densities of the GWB and detector noise (assumed to be the same for both detectors), and where we explicitly show the dependence of the overlap reduction function on the orientation angle $\phi_{\rm rot}$.
We can ignore the proportionality factor when calculating the expected squared-SNR, since we will ultimately be normalizing $\rho^2$ to have maximum value equal to one.

We now consider the following three cases of interest: (i) weak-signal limit (where $S\ll N$), (ii) intermediate-signal regime (where $S=N$), and (iii) the strong-signal limit (where $S\gg N$). 
For those three cases we have:
\be
\rho^2(\phi_{\rm rot})\big|_{\rm weak} = \frac{\gamma^2(\phi_{\rm rot}) S^2}{N^2}\,,
\qquad
\rho^2(\phi_{\rm rot})\big|_{\rm intermediate} = \frac{\gamma^2(\phi_{\rm rot})}{4+ \gamma^2(\phi_{\rm rot})}\,,
\qquad
\rho^2(\phi_{\rm rot})\big|_{\rm strong} = \frac{\gamma^2(\phi_{\rm rot})}{1+ \gamma^2(\phi_{\rm rot})}\,.
\ee
For all three cases, the minimum and maximum squared SNR values occur when $\gamma(\phi_{\rm rot})=0$ and $\gamma^2(\phi_{\rm rot})=1$, which correspond to $\phi_{\rm rot}=45^\circ$ and $0^\circ$, respectively.
Normalizing the above SNRs by their maximum values leads to
\be
\bar\rho^2(\phi_{\rm rot})\big|_{\rm weak} = \gamma^2(\phi_{\rm rot})\,,
\qquad
\bar\rho^2(\phi_{\rm rot})\big|_{\rm intermediate} = \frac{5\gamma^2(\phi_{\rm rot})}{4+ \gamma^2(\phi_{\rm rot})}\,,
\qquad
\bar\rho^2(\phi_{\rm rot})\big|_{\rm strong} = \frac{2 \gamma^2(\phi_{\rm rot})}{1+ \gamma^2(\phi_{\rm rot})}\,,
\ee
where we use an overbar on $\bar\rho$ to indicate that it has been normalized.
It is now a simple matter to calculate the ratio of the values of $\bar\rho(\phi_{\rm rot})$ at different values of $\phi_{\rm rot}$.
Finally, as we did in the main text, we will compare the values of $\bar\rho(\phi_{\rm rot})$ at $\phi_{\rm rot}=42.5^\circ$ and $0^\circ$ values.
Doing so we find:
\be
{\rm ratio}\big|_{\rm weak}
\approx \frac{1}{11}\,,
\qquad
{\rm ratio}\big|_{\rm intermediate}
\approx
\sqrt{\frac{5}{4}}\frac{1}{11}\approx\frac{1}{9.8}\,,
\qquad
{\rm ratio}\big|_{\rm strong}\approx \frac{\sqrt{2}}{11}\approx\frac{1}{7.7}\,.
\ee
So even in the strong-signal limit, there is a reduction in the expected SNR by a factor of approximately 8.

\end{document}